# ESTIMATING FUNCTIONS OF PROBABILITY DISTRIBUTIONS FROM A FINITE SET OF SAMPLES

## Part I: Bayes Estimators and the Shannon Entropy.

by

David H. Wolpert[1] and David R. Wolf[2]

1 -  The Santa Fe Institute, 1660 Old Pecos Trail, Suite A, Santa Fe, NM 87501.
(dhw@santafe.edu)

2 -  Image Analysis Section, DX Division, DX-13, MS P940, LANL, Los Alamos, NM 87545.
(wolf@lanl.gov)

Abstract: This paper is the first of two on the problem of estimating a function of a probability distribution from a finite set of samples of that distribution. In this paper a Bayesian analysis of this problem is presented, the optimal properties of the Bayes estimators are discussed, and as an example of the formalism, closed form expressions for the Bayes estimators for the moments of the Shannon entropy function are derived. Numerical results are presented that compare the Bayes estimator to the frequency-counts estimator for the Shannon entropy .

## 1. INTRODUCTION

Consider a system with m possible states and an associated m-vector of probabilities of those states, $\mathbf{p} = (p_i)$, $1 \leq i \leq m$, $(\Sigma_{i=1}^{m} p_i = 1)$. The system is repeatedly and independently sampled according to the distribution $\mathbf{p}$. Let the total number of samples be N and denote the associated vector of counts of states by $\mathbf{n} = (n_i)$, $1 \leq i \leq m$, $(\Sigma_{i=1}^{m} n_i = N)$. By definition, $\mathbf{n}$ is multinomially distributed.

In many cases what we are interested in is not $\mathbf{p}$ but some function of $\mathbf{p}$, $F(\mathbf{p})$. In these papers (this paper and Ref. 1) we are concerned with the problem of estimating some function $F(\mathbf{p})$ from the data $\mathbf{n}$. This problem is ubiquitous in physics, arising for example in dimension estimation and in estimating correlations from data. Some previous work on this issue (most closely related to the work of Ref. 2) appears in Refs. 3 4, 13, and 14.

In Sec. 2 of this paper we introduce the Bayes estimator for $F(\mathbf{p})$ given $\mathbf{n}$. In Sec. 3 we discuss the optimal properties of Bayes estimators and discuss their relation to conventional statistical techniques. Section 4 contains the central mathematical results needed to calculate Bayes estimators for $F(\mathbf{p})$. We then apply these results to the case where $F(\mathbf{p})$ is the Shannon entropy[5-7] $S(\mathbf{p}) = -\Sigma_i p_i \log(p_i)$. Section 5a contains a brief calculation showing that for small sample sizes there are significant differences between the Bayes and frequency-counts $(S(\mathbf{n}) = -\Sigma_i (n_i/N) \log(n_i/N))$ estimators for the Shannon entropy. In Sec. 5b we present graphs of the results of a numerical comparison of the Bayes and frequency-counts estimators for the Shannon entropy.

We note in passing that the intuitive notion of Shannon entropy as the "amount of missing information" is not usually considered meaningful if the information at hand consists of data $\mathbf{n}$ rather than the underlying distribution $\mathbf{p}$, since Shannon entropy is a function of $\mathbf{p}$ rather than of $\mathbf{n}$. In the sense that the Bayes estimator discussed in this paper is optimal, and produces a Shannon entropy value from information of the form $\mathbf{n}$, the Bayes estimator can be viewed as a way of defining the "amount of missing information" when the information at hand consists of a finite data set **n** rather than a full distribution **p**.

In the second paper in this series[1] Bayes estimators for several other functions $F(\mathbf{p})$ besides the Shannon entropy are presented. In a paper in preparation we will present the results of a classical sampling distribution analysis of the problem of estimating $F(\mathbf{p})$ from **n** for several $F(\mathbf{p})$ of interest[2].

## 2. BAYESIAN ESTIMATION OF F(p) FROM COUNTS

To estimate $F(\mathbf{p})$ from the data **n**, it is necessary to find the probability density function (pdf) $P(\mathbf{p} \mid \mathbf{n})$. First note that $P(\mathbf{n} \mid \mathbf{p}) = N! \Pi_{i=1}^{m} [p_i^{n_i} / n_i!]$. By Bayes' theorem the pdf $P(\mathbf{p} \mid \mathbf{n})$ is given by

$$P(\mathbf{p} \mid \mathbf{n}) = P(\mathbf{n} \mid \mathbf{p}) P(\mathbf{p}) / P(\mathbf{n}) \qquad (1)$$

where $P(\mathbf{n}) = \int d\mathbf{p} P(\mathbf{n} \mid \mathbf{p}) P(\mathbf{p})$, and where $P(\mathbf{p})$ has support only on the simplex $R \equiv \{ \mathbf{p} : p_i \geq 0 \, \forall \, i, \Sigma_i p_i = 1 \}$. $P(\mathbf{p} \mid \mathbf{n})$ is called the "posterior pdf", $P(\mathbf{n} \mid \mathbf{p})$ is called the "likelihood", and $P(\mathbf{p})$ is called the "prior pdf". Unless otherwise stated, integrals over **p** are understood to be definite integrals over the region extending from 0 to $\infty$ in each $p_i$. Note that because of cancellation, the constant $N! \Pi_{i=1}^{m} n_i!$ does not appear in $P(\mathbf{p} \mid \mathbf{n})$; we can simply write $P(\mathbf{p} \mid \mathbf{n}) \propto P(\mathbf{p}) \Pi_{i=1}^{m} p_i^{n_i}$ with the proportionality constant (dependent on **n** only) set by normalization.

The pdf of $F(\mathbf{p})$ given **n** is given in terms of $P(\mathbf{p} \mid \mathbf{n})$ by

$$P(F(\mathbf{p}) = f \mid \mathbf{n}) = \int d\mathbf{p} \; \delta(F(\mathbf{p}) - f) \; P(\mathbf{p} \mid \mathbf{n}) \qquad (2)$$

It is important to note that if what we know is **n**, and if what we wish to know is f, then it is the distribution in Eq. (2) and this one alone, which tells us what we want. Distributions which do not depend on a prior (e.g., likelihood-based quantities) do not share this property. (Of course, they also don't share the property of distributions like $P(F(\mathbf{p}) = f \mid \mathbf{n})$ that, as usually calculated, such distributions depend on an assumption for the prior. See Ref. 1.)

Rather than trying to find the density $P(F(\mathbf{p}) = f \mid \mathbf{n})$ directly, it is simpler to find the moments of $F^k(\mathbf{p})$ given this density. The $k^{th}$ moment of $F(\mathbf{p})$ given $\mathbf{n}$ is given by $\int df \, f^k \, P(F(\mathbf{p}) = f \mid \mathbf{n}) = \int d\mathbf{p} \, F^k(\mathbf{p}) P(\mathbf{p} \mid \mathbf{n})$, i.e. the $k^{th}$ moment of $F(\mathbf{p})$ given $\mathbf{n}$ is the posterior average of $F^k(\mathbf{p})$ according to the posterior distbution $P(\mathbf{p} \mid \mathbf{n})$. Define $F_k$ by

$$F_k \equiv \int d\mathbf{p} F^k(\mathbf{p}) P(\mathbf{p}) \, \Pi_{i=1}^{m} p_i^{n_i}. \qquad (3)$$

Using our formula for $P(\mathbf{p} \mid \mathbf{n})$ we see that the $k^{th}$ moment of $F(\mathbf{p})$ given $\mathbf{n}$ is given by $F_k / F_0$. We refer to this ratio as the "Bayes estimator with prior $P(\mathbf{p})$ for $F^k(\mathbf{p})$".

As an aside, note that the moments are useful things to know even when the distribution in question is non-gaussian, so that knowing such moments doesn't directly give us things like the full-width-half-maximum of the distribution.. In particular, the Bayes-estimators for $F^k(\mathbf{p})$ and $F^{2k}(\mathbf{p})$ can be used to find the standard deviation of $F^k(\mathbf{p})$. This in turn may be used with Chebyshev's inequality to bound the probability of deviation of $F^k(\mathbf{p})$ from the Bayes estimator for $F^k(\mathbf{p})$, even if the posterior is non-gaussian.

To proceed further it is necessary to make an assumption for the prior pdf $P(\mathbf{p})$; once this is done, $P(\mathbf{p} \mid \mathbf{n})$ and $F_k / F_0$ are uniquely determined. In the calculations to follow, $P(\mathbf{p})$ will be assumed to be a uniform prior, i.e. it will be assumed to have the form $P(\mathbf{p}) \propto \Delta(\mathbf{p})\Theta(\mathbf{p})$, where $\Theta(\mathbf{p}) \equiv \Pi_i \theta(p_i)$, ($\theta$ is the theta function $\theta(x) = 1$ for $x \geq 0$, 0 otherwise), $\Delta(\mathbf{p}) \equiv \delta(\Sigma_i p_i - 1)$, and the proportionality constant is set by the normalization condition $\int d\mathbf{p} \, P(\mathbf{p}) = 1$.

We emphasize that here we are using the uniform prior only for reasons of expository simplicity. In many problems the uniform prior is inappropriate and a different prior should be used. As an example of such a non-uniform prior, the entropic prior $P(\mathbf{p}) \propto e^{\alpha S}$, where S is the Shannon entropy and $\alpha$ is some constant, is related to the technique of Maximum Entropy[8]. (Also see footnote 1.) As another example, the Dirichlet prior, $P(\mathbf{p}) \propto \Sigma_{i=1}^m p_i^a$ for some constant a, has also been considered in some contexts[9]. It is also sometimes appropriate to assign to some states i a prior which does not allow $p_i$ to differ from zero[12].

In Ref. 1 we consider the extension of our results to a broader class of priors than those considered in this paper. In particular, both entropic and Dirichlet priors are discussed there. As a general rule, because there is no reason to believe that a Bayesian technique is optimal if the prior it uses is poorly chosen, we admonish the reader to choose each prior with careful attention to the problem at hand.

For simplicity of presentation define

$$I[F(\mathbf{p}), \mathbf{n}] \equiv \int d\mathbf{p}\, F(\mathbf{p})\, \Delta(\mathbf{p})\Theta(\mathbf{p})\Pi_{i=1}^m p_i^{n_i}. \qquad (4)$$

Note that $I[\cdot, \cdot]$ is a functional of its first argument and a function of its second argument. With this notation the Bayes estimator with uniform prior for $F^k(\mathbf{p})$ (i.e. $F_k / F_0$ with $P(\mathbf{p})$ uniform) is given by $I[F^k(\mathbf{p}), \mathbf{n}] / I[1, \mathbf{n}]$. (For non-uniform $P(\mathbf{p})$, $F_k / F_0$ is given by a different ratio of integrals.)

## 3. BAYES ESTIMATORS MINIMIZE MEAN-SQUARED ERROR

Before evaluating the integrals $I[F^k(\mathbf{p}), \mathbf{n}]$ we briefly discuss an optimality property of Bayes estimators and relate these estimators to some classical estimation techniques.

If the true probabilities are fixed to $\mathbf{p}$, then the mean-squared error when using an estimator $G(\mathbf{n})$ to estimate $F(\mathbf{p})$ is given by

$$\sum_{\mathbf{n}} P(\mathbf{n} \mid \mathbf{p}) \times (G(\mathbf{n}) - F(\mathbf{p}))^2. \tag{5}$$

For a given $\mathbf{p}$, (5) is minimized by choosing $G(\mathbf{n})$ independent of the $\mathbf{n}$: $G(\mathbf{n}) = F(\mathbf{p})$. More generally, when $\mathbf{p}$ is distributed according to $P(\mathbf{p})$, the mean-squared error is given by

$$\int d\mathbf{p} \, P(\mathbf{p}) \sum_{\mathbf{n}} P(\mathbf{n} \mid \mathbf{p}) \times (G(\mathbf{n}) - F(\mathbf{p}))^2. \tag{6}$$

As conventional, find the $G(\cdot)$ minimizing this expression by writing $G(\cdot) = G_0(\cdot) + \alpha\eta(\cdot)$, differentiating (6) with respect to $\alpha$, and evaluating the result at $\alpha = 0$. Doing this yields

$$\sum_{\mathbf{n}} \eta(\mathbf{n}) \int d\mathbf{p} \, P(\mathbf{n} \mid \mathbf{p}) \, P(\mathbf{p}) \times (G_0(\mathbf{n}) - F(\mathbf{p})) = 0. \tag{7}$$

Since this equality must hold for all $\eta(\cdot)$, for all $\mathbf{n}$

$$\int d\mathbf{p} \, P(\mathbf{n} \mid \mathbf{p}) \, P(\mathbf{p}) \times (G_0(\mathbf{n}) - F(\mathbf{p})) = 0. \tag{8}$$

Eqn. 8 is solved (assuming $\int d\mathbf{p} \, P(\mathbf{n} \mid \mathbf{p}) \, P(\mathbf{p}) \neq 0$) by

$$G_0(\mathbf{n}) = \int d\mathbf{p} \, P(\mathbf{n} \mid \mathbf{p}) \, P(\mathbf{p}) \, F(\mathbf{p}) \; / \; \int d\mathbf{p} \, P(\mathbf{n} \mid \mathbf{p}) \, P(\mathbf{p}) = F_1 / F_0. \tag{9}$$

Note that Eqn. 9 holds for any prior $P(\mathbf{p})$. Given the discussion in Sec. 2, Eqn. 9 shows that $G_0(\mathbf{n})$, the estimator having minimal mean-squared error from $F(\mathbf{p})$, is identical to the Bayes estimator for $F(\mathbf{p})$: $G_0(\mathbf{n}) = \int d\mathbf{p} P(\mathbf{p} \mid \mathbf{n}) F(\mathbf{p})$.

As an example consider the famous Laplace Sample Size Correction estimator, in which the underlying $p_i$ are estimated from counts $\mathbf{n}$ by $p_i = (n_i + 1) / (N + m)$. This estimator is precisely the Bayes estimator with uniform prior for $F(\mathbf{p}) = \mathbf{p}$ (see results in Ref. 1). Note that for small $n_i$ the Bayes estimator is especially different from the frequency count estimator $p_i = n_i / N$.

We note as an aside that when $F(.)$ is highly nonlinear and not one-to-one (e.g., when $F(.)$ is the Shannon entropy), one can not evaluate the Bayes estimate for $F(\mathbf{p})$ by calculating $F$ of the Bayes estimate for $\mathbf{p}$, i.e., by calculating $F((n_i + 1) / (N + m))$.) For these kinds of $F(.)$ one must take into account the probabilities of all $\mathbf{p}$'s to evaluate the Bayes estimator for $F(\mathbf{p})$. The set of the Bayes estimates for the individual $p_i$ simply does not contain sufficienct information to give the Bayes estimate for $F(\mathbf{p})$. (Never mind enough information to calculate error bars for that estimate.)

In general, one might not want to take the mean $f$ according to the pdf $P(F(\mathbf{p}) = f \mid \mathbf{n})$ to form an estimate for $F(\mathbf{p})$. For example, one might be interested in minimizing (the average of) $|G(\mathbf{n}) - F(\mathbf{p})|$ rather than $[G(\mathbf{n}) - F(\mathbf{p})]^2$, a goal which generically results in an estimate of the median of the pdf rather than its mean. As another example, in maximum-likelihood estimation, for the case where $F(\mathbf{p}) = \mathbf{p}$, one estimates $F(\mathbf{p})$ as the $\mathbf{p}$ that maximizes the likelihood $P(\mathbf{n} \mid \mathbf{p})$, rather than as a mean or a median. The maximum-likelihood estimate corresponds to finding the mode of $P(F(\mathbf{p}) = f \mid \mathbf{n})$ (assuming the prior over f is uniform). (When $F(\mathbf{p}) = \mathbf{p}$ the result is the frequency counts estimate, $p_i = n_i / N$.) As yet another example, it might be of interest to minimize something other than a functional of the error $G(\mathbf{n}) - F(\mathbf{p})$. An instance of this appears in footnote 2, which discusses minimizing the mean-squared bias to find what might be called a "Bayes minimum-bias estimator". Finally, note that the classical sampling distribution problem (arising in hypothesis testing) of finding the distribution of counts $\mathbf{n}$ given a known value f of $F(\mathbf{p})$, i.e. of finding $P(\mathbf{n} \mid F(\mathbf{p}) = f)$, may be handled using the techniques developed in this paper for calculating the posterior $P(F(\mathbf{p}) = f \mid \mathbf{n})$. We will discuss this issue in a later paper[2].

## 4. CALCULATION OF THE BAYES ESTIMATOR FOR SHANNON ENTROPY.

As was shown in Sec. 2, finding the Bayes estimator with uniform prior for $F^k(\mathbf{p})$ reduces to evaluating integrals of the form $I[F^k(\mathbf{p}) , \mathbf{n}]$. This section contains the central techniques for calculating these integrals and uses them to calculate the Bayes estimator for the Shannon entropy.

Readers interested only in the Shannon entropy results may skip directly to Sec. 4e. In Sec. 4a we derive an important result that allows integrals like $I[\cdot, \cdot]$ to be recast as Laplace convolution-products. In section 4b we outline the general procedure, based on the results of Sec. 4a, for calculating the moments of $F(\mathbf{p})$. In the remaining subsections we apply the procedure of Sec. 4b to the case in which $F(\mathbf{p})$ is the Shannon entropy. Section 4c contains a calculation of $F_0 = I[1 , \mathbf{n}]$. In Sec. 4d we present a calculation for those integrals which, along with $F_0$, give the Bayes esti-mator of the Shannon entropy.

### 4a.  CONVOLUTION FORM OF THE INTEGRALS

In this subsection two important results are given.  First, in Thm. 1 it is shown that if a function $H(\mathbf{p})$ factors as $H(\mathbf{p}) = \Pi_{i=1}^{m} h_i(p_i)$, then the general form of the integral $\int d\mathbf{p}\, H(\mathbf{p})\Delta(\mathbf{p})\Theta(\mathbf{p})$ is that of a convolution product of m terms (recall that m is the number of possible outcomes of the process under observation).  Second, Laplace's convolution theorem is given.

Define the Laplace convolution operator $\otimes$ by $(f \otimes g)(\tau) \equiv \int_0^{\tau} dx\, f(x)\, g(\tau\text{-}x)$.

<u>Theorem 1.</u>    If  $H(\mathbf{p}) = \Pi_{i=1}^{m} h_i(p_i)$ then $\int d\mathbf{p}\, \Delta(\mathbf{p})\Theta(\mathbf{p})\, H(\mathbf{p}) = \left. (\otimes_{i=1}^{m} h_i(p_i))(\tau)\right|_{\tau=1}$.

<u>Proof:</u>   The $p_i$ may not be independently integrated since the constraint $\Sigma_{i=1}^{m} p_i = 1$ exists.  This constraint is reflected in the explicit definition of the integral,

$$\int d\mathbf{p}\, \Delta(\mathbf{p})\Theta(\mathbf{p})\, H(\mathbf{p}) = \int_0^{\infty} dp_1 \int_0^{\infty} dp_2 \dots \{h_1(p_1) \times \dots \times h_m(p_m)\} \times \delta(1 - \Sigma_{i=1}^{m} p_i)$$

$$= \int_0^1 dp_1 h_1(p_1) \int_0^{1-p_1} dp_2 h_2(p_2) \dots$$

$$\dots \int_0^{1-(p_1+\dots+p_{m-2})} dp_{m-1} h_{m-1}(p_{m-1}) h_m(1-(p_1+\dots+p_{m-1}))$$

Define the m variables $\tau_k$, k = 1, ... , m, recursively by $\tau_1 \equiv \Sigma_{i=1}^{m} p_i$ and $\tau_k \equiv \tau_{k-1} - p_{k-1}$. Since $\tau_k = \tau_1 - \Sigma_{i=1}^{k-1} p_i$, our integral may be rewritten as

$$\int d\mathbf{p}\, \Delta(\mathbf{p})\Theta(\mathbf{p})\, H(\mathbf{p}) =$$

$$\left. \int_0^{\tau_1} dp_1 h_1(p_1) \int_0^{\tau_2} dp_2 h_2(p_2) \dots \int_0^{\tau_{m-1}} dp_{m-1} h_{m-1}(p_{m-1}) h_m(\tau_{m-1} - p_{m-1}) \right|_{\tau_1 = 1}.$$

Now, with the definition of the convolution the integral can be rewritten as

$$\int d\mathbf{p}\ \Delta(\mathbf{p})\Theta(\mathbf{p})\ H(\mathbf{p})\ =$$

$$\int_0^{\tau_1} dp_1 h_1(p_1)\dots \int_0^{\tau_{m-2}} dp_{m-2} h_{m-2}(p_{m-2})\ (h_{m-1}\otimes h_{m-2})\ (\tau_{m-2}-p_{m-2})\Big|_{\tau_1=1}.$$

Since the convolution operator is both commutative and associative, we can repeat this procedure and write the integral above with obvious notation as

$$\int d\mathbf{p}\ \Delta(\mathbf{p})\Theta(\mathbf{p})\ H(\mathbf{p})\ =\ (\otimes_{i=1}^m h_i(p_i))\,(\tau)\Big|_{\tau=1}.$$

QED.

Theorem 2 is the Laplace convolution theorem and is stated for completeness only. The proof may be found in Ref. 10. Define the Laplace transform operator L by $L[\,h\,](s) = \int_0^\infty h(t)e^{-st}dt$.

Theorem 2.   If $L[h_i(p_i)]$ exists for $i = 1, \dots, m$, then $L[\otimes_{i=1}^m h_i(p_i)] = \Pi_{i=1}^m L[h_i(p_i)]$.

## 4b.   OUTLINE OF GENERAL PROCEDURE

Theorems 1 and 2 allow the calculation of integrals $I[F^k(\mathbf{p})\,,\,\mathbf{n}]$ for functions of the form $F(\mathbf{p}) = \Sigma_{i=1}^k \Pi_{j=1}^m h_{ij}(p_j)$, which we call "factorable". Here we briefly summarize the procedure to be used.

   i) For each $h_{ij}(p_j)$, calculate the Laplace transform of $h_{ij}(p_j) \times p_j^{n_j}$.

   ii) Calculate $\Sigma_{i=1}^k \Pi_{j=1}^m L[h_{ij}(p_j) \times p_j^{n_j}]$.

   iii) Take the inverse Laplace transform of the term calculated in (ii).

As an example, let $F(\mathbf{p}) = S(\mathbf{p}) = -\Sigma_{i=1}^{m} p_i \log(p_i)$. All powers of $S(\mathbf{p})$ are factorable terms. Therefore, the procedure outlined above may be used to find the Bayes estimators with uniform prior for any power of $S(\mathbf{p})$, as shown in detail in the remainder of this section.

## 4c.  CALCULATION OF I[1 , n]

In the next theorem the Laplace transform is used in concert with Thms. 1 and 2 to calculate the normalization constant $F_0 = I[1 , \mathbf{n}]$. Defining the Gamma function $\Gamma(z)$ by

$$\Gamma(z) = \int_0^\infty t^{z-1} e^{-t} dt \text{ for Re } (z) > -1 \text{ we have}$$

Theorem 3.    If $Re(n_i) > -1 \; \forall \; i = 1, ..., m$, then $I[1 , \mathbf{n}] = \dfrac{\Pi_{i=1}^{m} \Gamma(n_i + 1)}{\Gamma[N + m]}$.

Proof:  For the integral $I[1 , \mathbf{n}] = \int d\mathbf{p} \; \Delta(\mathbf{p}) \Theta(\mathbf{p}) \Pi_{i=1}^{m} p_i^{n_i}$, the $h_i(p_i)$ of Thm. 1 are given by $h_i(p_i) = p_i^{n_i}$. Since

$$L[p^n](s) = \frac{\Gamma(n+1)}{s^{n+1}} \quad \text{for } n > -1,$$

we have by Thms. 1 and 2

$$I[1 , \mathbf{n}] = L^{-1}[\Pi_{i=1}^{m} L[p^{n_i}](s)] \Big|_{\tau=1} = L^{-1}\left[\Pi_{i=1}^{m} \frac{\Gamma(n_i + 1)}{s^{n_i}}\right](\tau) \Big|_{\tau=1}$$

$$= \frac{\Pi_{i=1}^{m} \Gamma(n_i + 1)}{\Gamma(N + m)}.$$

<div align="right">QED.</div>

## 4d.  CALCULATION OF I[ $p_1^{q_1} \log^{r_1}(p_1) \ldots p_m^{q_m} \log^{r_m}(p_m)$ , n]

As mentioned in Sec. 4b, since $S(\mathbf{p}) = -\Sigma_i p_i \log(p_i)$, powers of $S(\mathbf{p})$ are sums of terms each of which have the form $p_1^{q_1} \log^{r_1}(p_1) \ldots p_m^{q_m} \log^{r_m}(p_m)$. Thus, to find the Bayes estimators for arbitrary powers of the Shannon entropy, expressions of the form

$$I[p_1^{q_1} \log^{r_1}(p_1) \ldots p_m^{q_m} \log^{r_m}(p_m), \mathbf{n}]$$

must be calculated. Using the fact that $\partial_n^r \, p^n = p^n \log^r(p)$, we immediately have

Theorem 4.  For $\mathrm{Re}\,(n_i) > -1 \; \forall\, i$,

$$I[\log^{r_1}(p_1) \ldots \log^{r_m}(p_m), \mathbf{n}] = \partial_{n_1}^{r_1} \ldots \partial_{n_m}^{r_m} I[1, \mathbf{n}].$$

The justification for the needed interchange of derivative and integral is given in App. C of Ref. 1. In using Thm. 4, note that since $N \equiv \Sigma_{i=1}^m n_i$ we have $\partial_i N = 1$.

Make the definitions $\Phi^{(n)}(z) \equiv \Psi^{(n-1)}(z)$, and $\Delta\Phi^{(n)}(z_1, z_2) \equiv \Phi^{(n)}(z_1) - \Phi^{(n)}(z_2)$, where $\Psi^{(n)}(z)$ is the polygamma function $\Psi^{(n)}(z) = \partial_z^{n+1} \log(\Gamma(z))$ [11]. This definition of $\Phi$ is made to facilitate the clean presentation of results; $\Phi^{(n)}(z) = \partial_z^n \log(\Gamma(z))$.

Thms. 5 and 6 apply Thm. 4 to the calculation of the integral $I[\log^{r_1}(p_1) \ldots \log^{r_m}(p_m), \mathbf{n}]$ for some special cases.

Theorem 5.      For $\mathrm{Re}\,(n_i) > -1 \; \forall\, i$,

$$I[\log(p_u), \mathbf{n}] = \{\Pi_i \, \Gamma(n_i+1)\} / \Gamma(N+m) \times \Delta\Phi^{(1)}(n_u+1, N+m).$$

Proof:      $I[\log(p_u), \mathbf{n}] = \partial_{n_u} I[1, \mathbf{n}]$                    (by Thm. 4)

Substituting the result from Thm. 3 for $I[1, \mathbf{n}]$ above we find

$$= \partial_{n_u} \frac{\Pi_{i=1}^m \Gamma(n_i+1)}{\Gamma(N+m)} = \Pi_{i \neq u} \Gamma(n_i+1) \partial_{n_u} \frac{\Gamma(n_u+1)}{\Gamma(N+m)}$$

$$= \Pi_{i \neq u} \Gamma(n_i + 1) \frac{\Gamma(n_u + 1)}{\Gamma(N + m)} \times \Delta \Phi^{(1)}(n_u + 1, N + m) \qquad \text{(by Def. of } \Phi)$$

$$= \frac{\Pi_{i=1}^{m} \Gamma(n_i + 1)}{\Gamma(N + m)} \times \Delta \Phi^{(1)}(n_u + 1, N + m). \qquad \text{QED.}$$

<u>Theorem 6.</u>  For Re $(n_i) > -1 \ \forall \, i$,

    i)    $I[ \log(p_u) \log(p_v) , \mathbf{n}] = \{ \Pi_i \Gamma(n_i + 1) \} / \Gamma(N + m)$

            $\times \{ \Delta \Phi^{(1)}(n_u + 1 , N + m) \Delta \Phi^{(1)}(n_v + 1, N + m) - \Phi^{(2)}(N + m) \}. \quad u \neq v .$

    ii)   $I[ \log^2(p_u) , \mathbf{n}] = \{ \Pi_i \Gamma(n_i + 1) \} / \Gamma(N + m)$

            $\times \{ \Delta \Phi^{(1)}(n_u + 1 , N + m)^2 + \Delta \Phi^{(2)}(n_u + 1 , N + m) \}.$

<u>Proof:</u>   Similar to proof of Thm. 5.

## 4e.  THE BAYES ESTIMATORS FOR MOMENTS OF THE SHANNON ENTROPY.

In this subsection the results for the Bayes estimators with uniform prior for the first two powers of $S(\mathbf{p})$ are given, i.e. $S_1 / S_0$ and $S_2 / S_0$. Refer to Secs. 4a-d for the calculations used here, and to Sec. 4d for the definitions of the functions $\Phi^{(n)}(z)$ and $\Delta \Phi^{(n)}(z_1, z_2)$.

<u>Theorem 7.</u>    For Re $(n_i) > -1 \ \forall \, i$, $S_1 / S_0 = -\Sigma_i \dfrac{n_i + 1}{N + m} \Delta \Phi^{(1)}(n_i + 2 , N + m + 1).$

<u>Proof:</u>    $S_1 / S_0 = I[- \Sigma_i \ p_i \log(p_i) , \mathbf{n}] / \ I[1 , \mathbf{n}]$

           $= -\Sigma_i \ I[\log(p_i) , \mathbf{n} + \mathbf{e}_i] / \ I[1 , \mathbf{n}] \qquad \text{(by Def. of } I[\cdot , \cdot])$

           $= -\Sigma_i \ \partial_{n_i} I[1 , \mathbf{n} + \mathbf{e}_i] / \ I[1 , \mathbf{n}] \qquad \text{(by Thm. 4)}$

$$= -\Sigma_i \frac{n_i + 1}{N + m} \, \Delta\Phi^{(1)}(n_i + 2 \, , \, N + m + 1) \qquad \text{(by Thms. 5, 3).}$$

QED.

<u>Theorem 8.</u>   For $\text{Re}(n_i) > -1 \;\forall\, i$, $S_2 / S_0 =$

$$\Sigma_{i \ne j} \frac{(n_i + 1)\,(n_j + 1)}{(N + m)\,(N + m + 1)} \, \{ \, \Delta\Phi^{(1)}(n_i + 2 \, , \, N + m + 2) \; \Delta\Phi^{(1)}(n_j + 1 \, , \, N + m + 2) - \Phi^{(2)}(N + m + 2) \, \}$$

$$+ \; \Sigma_i \frac{(n_i + 1)\,(n_i + 2)}{(N + m)\,(N + m + 1)} \times \{ \, \Delta\Phi^{(1)}(n_i + 3 \, , \, N + m + 2)^2 + \Delta\Phi^{(2)}(n_i + 3 \, , \, N + m + 2) \, \}.$$

<u>Proof:</u>   Similar to proof of Thm. 7.

In a manner similar to the calculation of $S_1$ and $S_2$, all higher moments of $S(\mathbf{p})$ are calculable via differentiation (since $\partial_z \Phi^{(n)}(z) = \Phi^{(n+1)}(z)$). Note that when no data have been observed, i.e. $\mathbf{n} = 0$, the estimator for $S_1 / S_0$ is simply $-\Delta\Phi^{(2)}(2, m + 1) = \Sigma_{i=2}^{m} i^{-1}$. It should also be noted that as $N \to \infty$ the Bayes estimator $S_1 / S_0 \to -\Sigma_i (n_i / N) \log(n_i / N)$, i.e. it asymptotically becomes the frequency-counts estimator[11].

## 5. BAYES ESTIMATORS VS. FREQUENCY-COUNTS ESTIMATORS.

In this section we compare the Bayes estimator (see Thm. 7) and the frequency-counts estimator for the entropy in two ways. First, in Sec. 5a an explicit calculation of the two estimators is made for two specific cases when a small number of counts are observed in two bins ($m = 2$). This simple calculation points out that for small $N$ there are significant differences in the values of the two estimators. Second, in Sec. 5b the two estimators are graphically compared for a range of sample sizes and true underlying distributions.

## 5a. SMALL N

For small N, the Bayes estimate $S_1 / S_0$ can differ considerably from the estimate one would make using the frequency-counts estimator, $S(\mathbf{n}) = -\Sigma_{i=1}^{m} (n_i / N) \log (n_i / N)$. This is illustrated by the following pair of examples:

Example 1: Assume two possible events (m = 2). Let $n_1 = 0$, and $n_2 = 2$. $S_1 / S_0 = .458$. The entropy estimate obtained using the frequency-counts estimator is 0.

Example 2: Again, have m = 2. Assume that $n_1 = 1$ and $n_2 = 4$. $S_1 / S_0 = .533$. The entropy estimate obtained using the frequency-counts estimator is 0.500.

Note that there are "edge effects" in using $S_1 / S_0$ as the estimate for the entropy. If the true $\mathbf{p}$ is uniform ($p_i = m^{-1} \forall i$), then $S(\mathbf{p})$ is maximal and always exceeds $S_1 / S_0$, no matter what the observed $\mathbf{n}$ are. This is because the estimate $S_1 / S_0$ takes into account all possible $\mathbf{p}$ which might have generated the observed n, including all those with a smaller entropy than the true (maximal) entropy. In a similar fashion, if the true $S(\mathbf{p})$ is minimal then it is always exceeded by $S_1 / S_0$, regardless of the value of the observed $\mathbf{n}$.

## 5b. GRAPHICAL RESULTS OF NUMERICAL COMPARISONS.

The graphs appearing in Figs. 1-5 depict several comparisons of the Bayes and frequency-counts estimators for entropy. In all cases the solid line represents the Bayes estimator, the dash-dot line represents the frequency-counts estimator, and the dotted line represents the true value of the entropy, where applicable. Figure 6 depicts the pdf of the Bayes estimator for a fixed ratio of counts as the number of counts increases. The graphs are the result of exact numerical computations of the various quantities represented.

Figure 1 explicitly demonstrates the result of Sec. 2 of this paper for the Shannon entropy with m = 2. Recall that this section shows that the Bayes estimator is the minimal mean-squared error estimator. As is immediately seen in Fig. 1, for all N the Bayes estimator has a smaller mean-squared error than the frequency-counts estimator, where the mean-squared error for an estimator $S(\mathbf{n})$ is given by

$$\int d\mathbf{p} \ P(\mathbf{p}) \ \Sigma_{\mathbf{n}} \ P(\mathbf{n} \mid \mathbf{p}) \ \{S(\mathbf{n}) - S(\mathbf{p})\}^2. \qquad (10)$$

The curves were generated with $P(\mathbf{p})$ uniform. The Bayes estimator is that of Thm. 7 which assumes this uniform $P(\mathbf{p})$.

Figure 2 depicts the average over $\mathbf{p}$ of the sample variance, that is

$$\int d\mathbf{p} \ P(\mathbf{p}) \ \Sigma_{\mathbf{n}} \ P(\mathbf{n} \mid \mathbf{p}) \ \{S(\mathbf{n}) - \Sigma_{\mathbf{n}'} P(\mathbf{n}' \mid \mathbf{p}) \ S(\mathbf{n}')\}^2. \qquad (11)$$

This figure shows how, for a particular sample size N, the estimators deviate from their sample averages. It is immediately seen that the Bayes estimator has a smaller sample variance. (This is in agreement with the conservative "edge effects" behavior of the Bayes estimator which was mentioned in the preceding subsection.) This result is useful for understanding Figs. 3 and 4.

Figure 3 shows the sample averages of the estimators as functions of the sample size N for various values of the true $\mathbf{p}$, that is

$$\Sigma_{\mathbf{n}} \ P(\mathbf{n} \mid \mathbf{p}) \ S(\mathbf{n}). \qquad (12)$$

Figure 4 shows the same sample averages of the estimators, but now as functions of the true $\mathbf{p}$ for various values of the sample size N.

It is of interest to note that the sample average of the frequency-counts estimator actually comes closest to the true entropy for a range of $\mathbf{p}$ values and sufficiently large N (see Figs. 3d-f and 4d-f). To see how this is possible in light of the fact that the Bayes estimator has lower mean-squared error, first note that

$$\int d\mathbf{p}\, P(\mathbf{p})\, \Sigma_{\mathbf{n}}\, P(\mathbf{n} \mid \mathbf{p}) \{S(\mathbf{n}) - S(\mathbf{p})\}^2 \ = \quad\quad\quad\quad\quad (13)$$

$$\int d\mathbf{p}\, P(\mathbf{p})\, \Sigma_{\mathbf{n}}\, P(\mathbf{n} \mid \mathbf{p}) \{S(\mathbf{n}) - \Sigma_{\mathbf{n}'} S(\mathbf{n}')\, P(\mathbf{n}' \mid \mathbf{p})\}^2 \ + \int d\mathbf{p}\, P(\mathbf{p}) \{\Sigma_{\mathbf{n}}\, P(\mathbf{n} \mid \mathbf{p})\, S(\mathbf{n}) - S(\mathbf{p})\}^2 \,,$$

i.e. the mean-squared error is the sum of the mean sample variance and the mean-squared bias. The left hand side of Eqn. 13 is depicted in Fig. 1. The first integral on the right hand side is depicted in Fig. 2. The integrand of the last integral on the right hand side (excluding $P(\mathbf{p})$) appears in Fig. 3 as the square of the difference between the curve for the estimator, and the true value being estimated. This quantity favors the frequency-counts estimator for some values of $\mathbf{p}$ for sufficiently large N; however the first integral on the right more than compensates to give a result favoring the Bayes estimator.

Figure 5 depicts the sample averages of the estimators' deviations from true as a function of $\mathbf{p}$ for various sample sizes N,

$$\Sigma_{\mathbf{n}}\, P(\mathbf{n} \mid \mathbf{p}) \{S(\mathbf{n}) - S(\mathbf{p})\}^2. \quad\quad\quad\quad (14)$$

The integral of the expression in (14) multiplied by the density $P(\mathbf{p})$ (here uniform), depicted for various N, is shown in Fig. 1.

Finally, Fig. 6 shows the convergence of the pdf $P(s \mid \mathbf{n})$ given by

$$P(S(\mathbf{p}) = s \mid \mathbf{n}) \equiv \int d\mathbf{p}\, \delta(S(\mathbf{p}) - s)\, P(\mathbf{p} \mid \mathbf{n}) \quad\quad\quad (15)$$

for a fixed ratio (1 : 15) of observed counts $n_1 : n_2$, as the overall number of counts $N = n_1 + n_2$ increases. Note the increasing density placed upon the true entropy as the counts N increase. Note that the average of s according to this density $P(s \mid \mathbf{n})$, i.e. $\int ds\, s\, P(s \mid \mathbf{n})$, is the Bayes estimator for $S(\mathbf{p})$ given the observations $\mathbf{n}$. As mentioned previously, of all estimators, its squared error averaged over both $\mathbf{p}$ and $\mathbf{n}$ is minimal.

## ACKNOWLEDGEMENTS

We would like to thank James Theiler for supplying the code xyplot used to produce the figures. We would also like to thank the Center for Nonlinear Studies and Theoretical Division, both of LANL, for their partial support during part of this work. This work was supported in part by the United States Department of Energy under contract number W-7405-ENG-36. For part of this work David H. Wolpert was supported by the Santa Fe Institute and by National Library of Medicine grant number NLM F37 LM00011.

**FOOTNOTES**

[1] If one wishes to estimate **p** by finding the mode of $P(\mathbf{p} \mid \mathbf{n})$, then the entropic prior leads immediately to the technique of MaxEnt in the case where the data are not a finite vector of counts **n** but rather is some expectation value, $b = \Sigma_i\, p_i\, B(p_i)$. This follows since $P(\Sigma_i\, p_i\, B(p_i) = b \mid \mathbf{p})$, considered as a function of **p** with b fixed, is everywhere either 1 or 0. As a result, by Bayes' theorem, for a prior of the form $e^{\alpha S(\mathbf{p})}$, finding the mode of $P(\mathbf{p} \mid \Sigma_i\, p_i\, B(p_i) = b)$ is equivalent to maximizing $S(\mathbf{p})$ subject to $\Sigma_i\, p_i\, B(p_i) = b$.

2 Note that what we have shown here is that $F_1 / F_0$ has the least mean-squared error from the true $F(\mathbf{p})$, on average. This does not imply that it is the estimator of the entropy which is least *biased* on average. To find the least average bias estimator, one searches for the $G(\cdot)$ minimizing

$$\int d\mathbf{p}\, P(\mathbf{p}) \,[\, \Sigma_{\mathbf{n}} \,[\, P(\mathbf{n} \mid \mathbf{p}) \times G(\mathbf{n}) \,] - F(\mathbf{p}) \,]^2.$$

It is more complicated to find the $G(\cdot)$ minimizing this expression than it is to find the $G(\cdot)$ minimizing the expression in Sec. 2. Setting up the problem analogously to Sec. 2, we get

$$0 = \int d\mathbf{p}\, P(\mathbf{p}) \,[\, \Sigma_{\mathbf{n}} \,[\, P(\mathbf{n} \mid \mathbf{p})\, G(\mathbf{n}) \,] - F(\mathbf{p}) \,] \times [\, \Sigma_{\mathbf{n}}\, P(\mathbf{n} \mid \mathbf{p}) \times \eta(\mathbf{n}) \,].$$

As usual, for this to hold for all $\eta(\cdot)$ means that it must hold for $\eta(\cdot)$ a Kronecker delta function centered about any particular pattern **n**. Let k be any such pattern. We have

$$0 = \int d\mathbf{p}\, P(\mathbf{p})\, P(\mathbf{k} \mid \mathbf{p}) \,[\, \Sigma_{\mathbf{n}}\, P(\mathbf{n} \mid \mathbf{p})\, G(\mathbf{n})\ -\ F(\mathbf{p}) \,],$$

$$\sum_{\mathbf{n}} G(\mathbf{n}) \int d\mathbf{p} \, P(\mathbf{p}) \, P(\mathbf{n} \mid \mathbf{p}) \, P(\mathbf{k} \mid \mathbf{p}) = \int d\mathbf{p} \, P(\mathbf{p}) \, P(\mathbf{k} \mid \mathbf{p}) \, F(\mathbf{p}).$$

We can evaluate both of the integrals for any $\mathbf{n}$ and $\mathbf{k}$ (see Sec. 3). What we then have is a set of simultaneous equations, one for each value of $\mathbf{k}$. Each of the equations is of the form "linear combination of the $G(\mathbf{n})$ equals constant". (The linear combination being over all possible $\mathbf{n}$.) To solve for the least average biased $G(\cdot)$, we must solve this set of simultaneous equations.

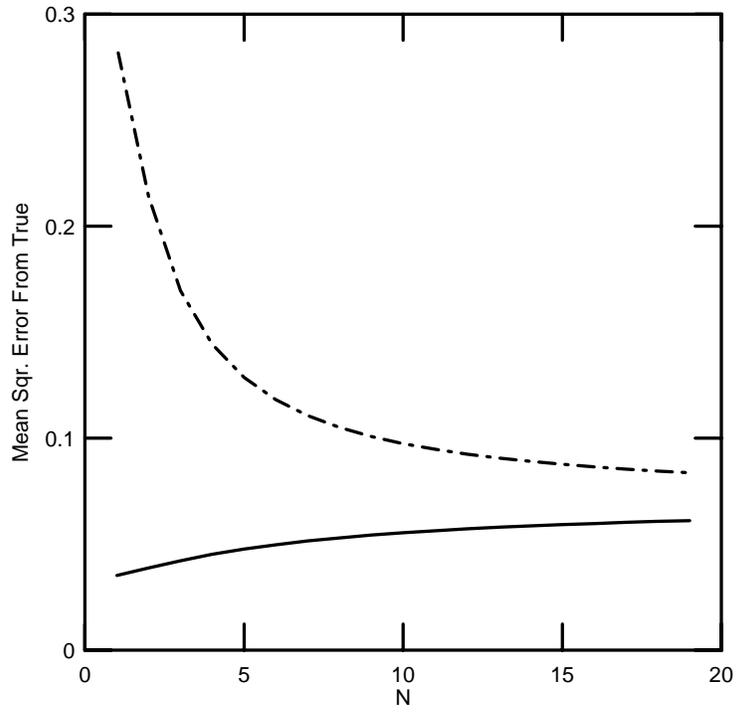

Fig. 1

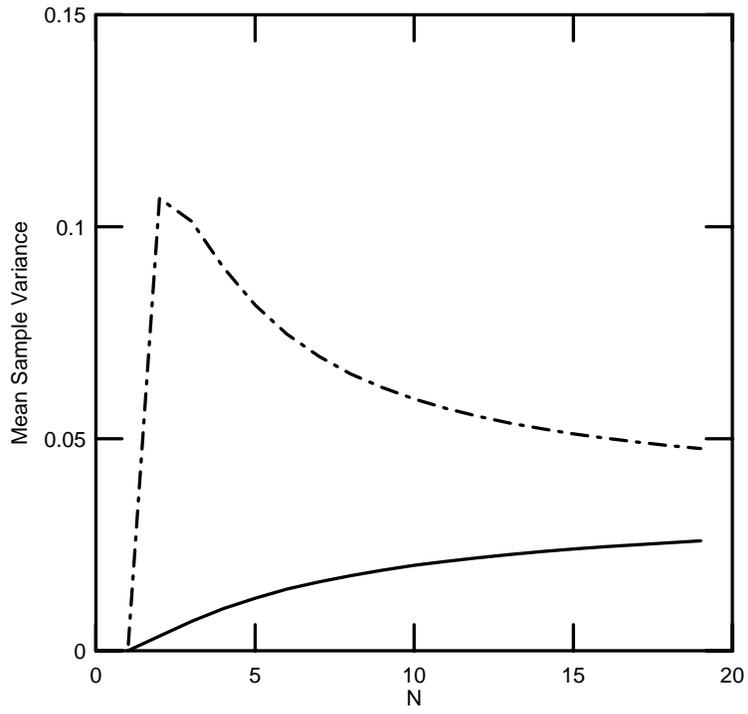

Fig. 2

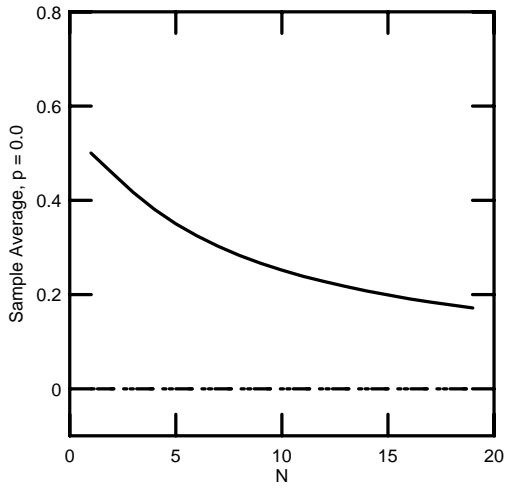

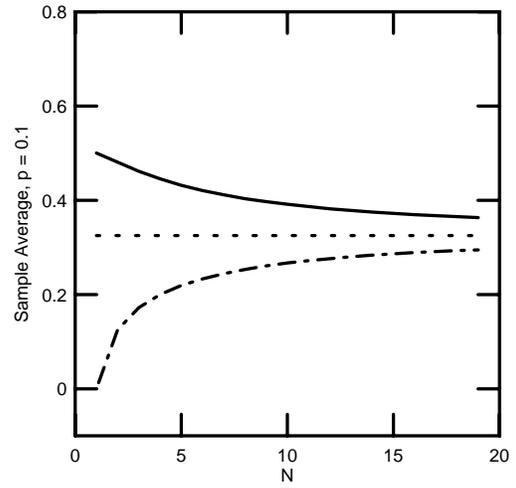

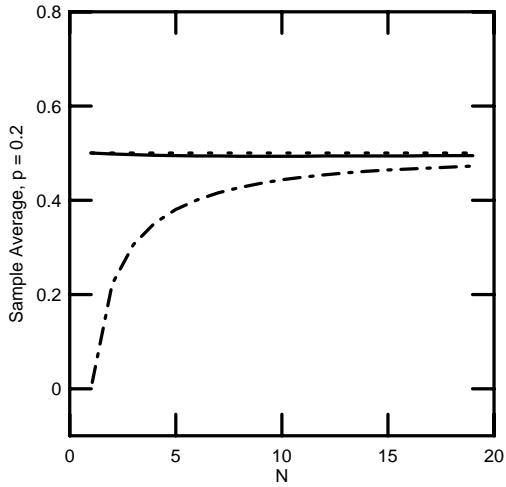

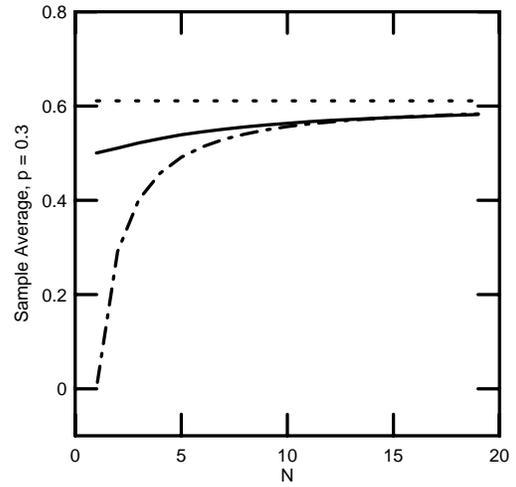

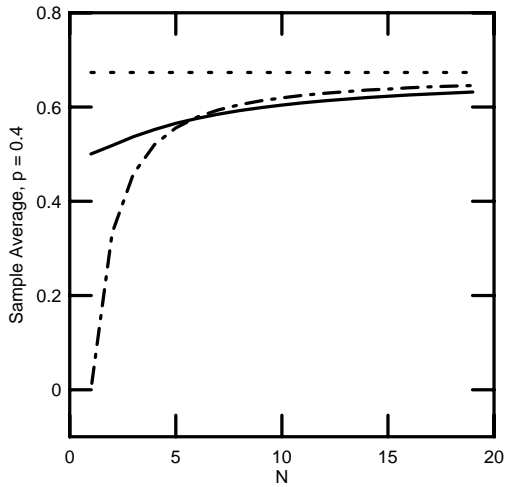

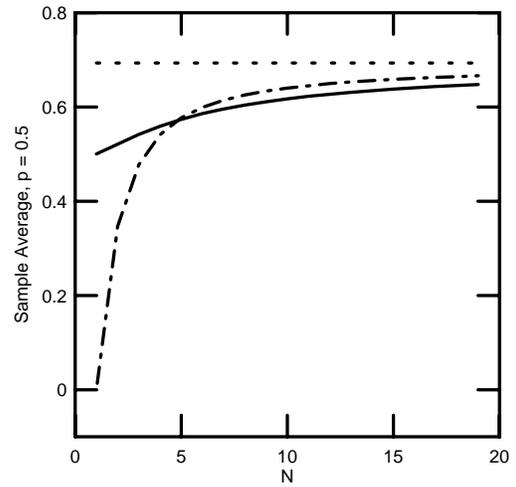

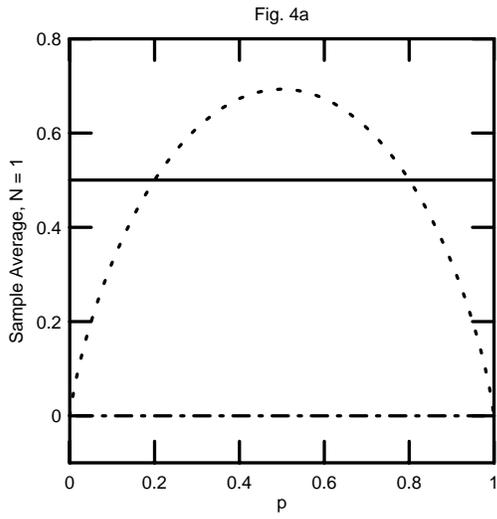

Fig. 4a

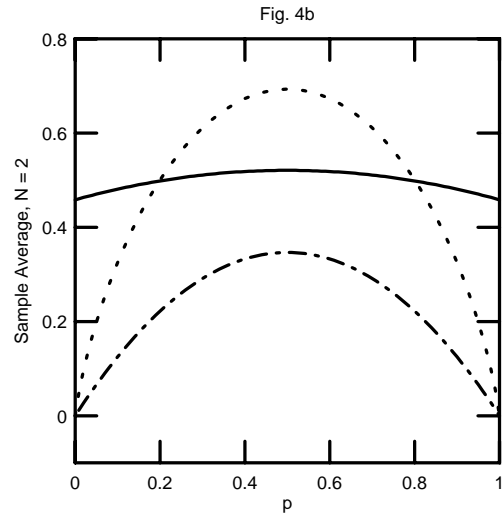

Fig. 4b

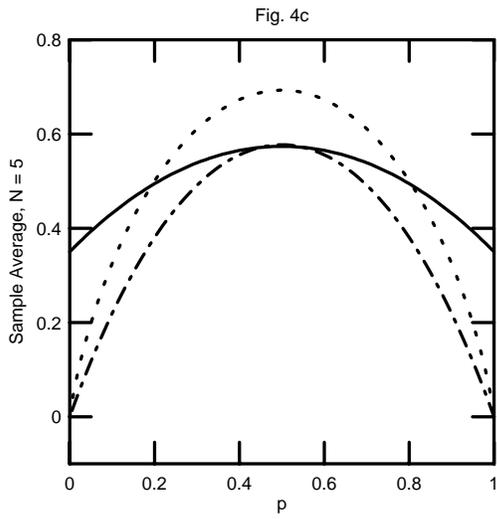

Fig. 4c

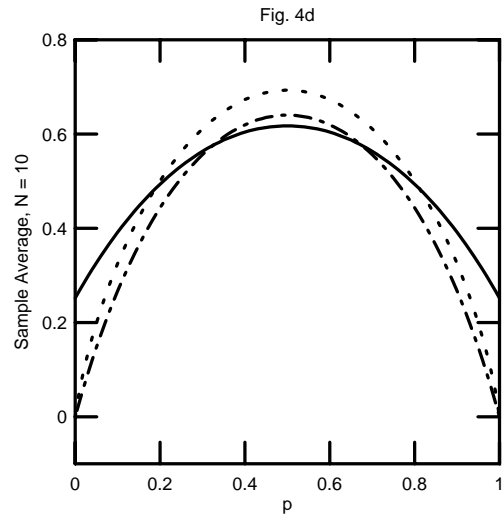

Fig. 4d

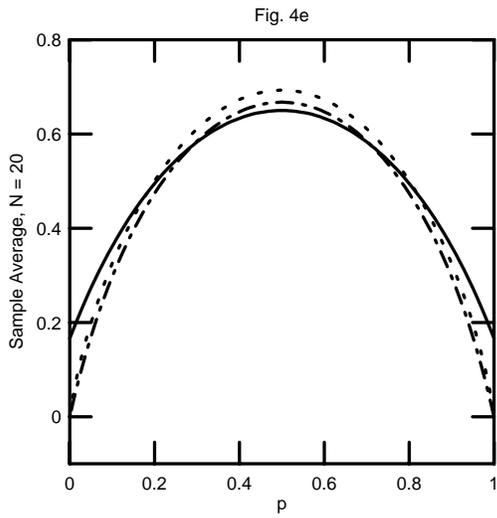

Fig. 4e

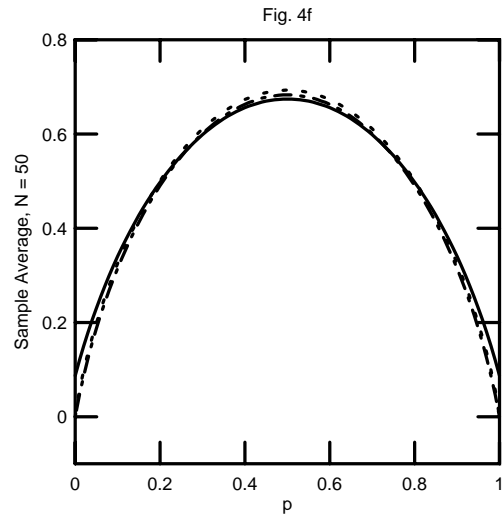

Fig. 4f

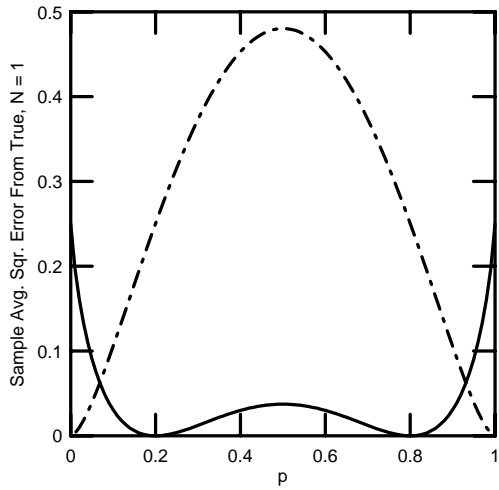

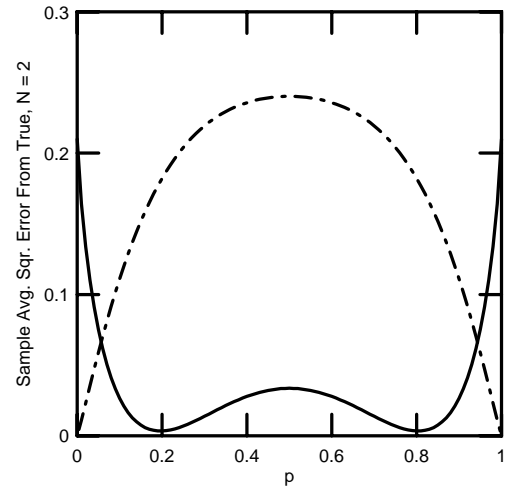

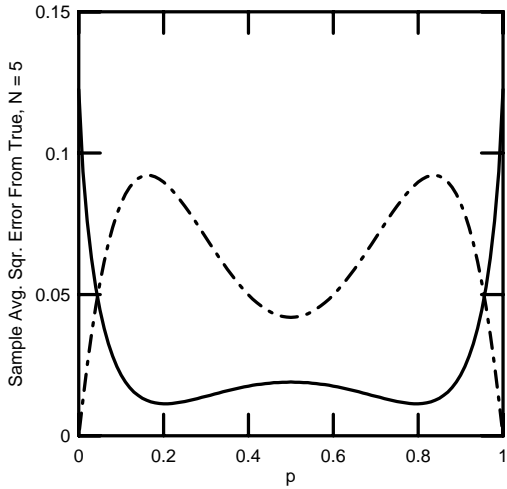

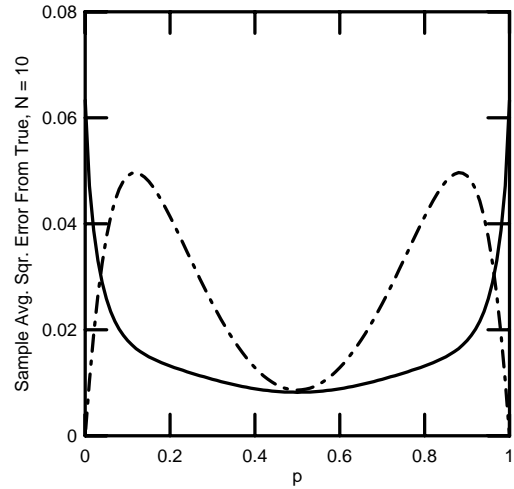

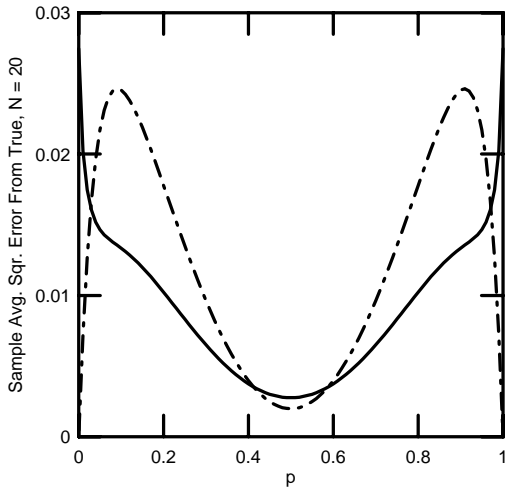

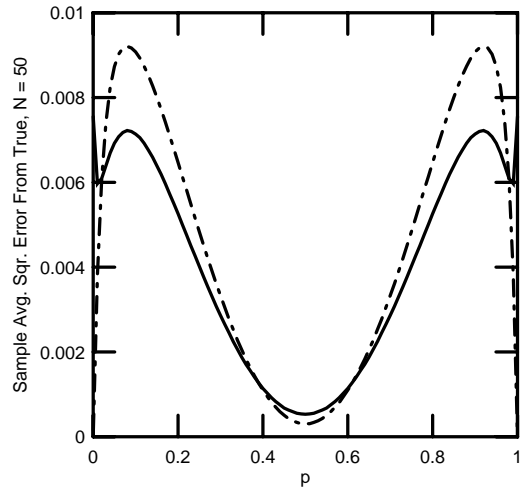

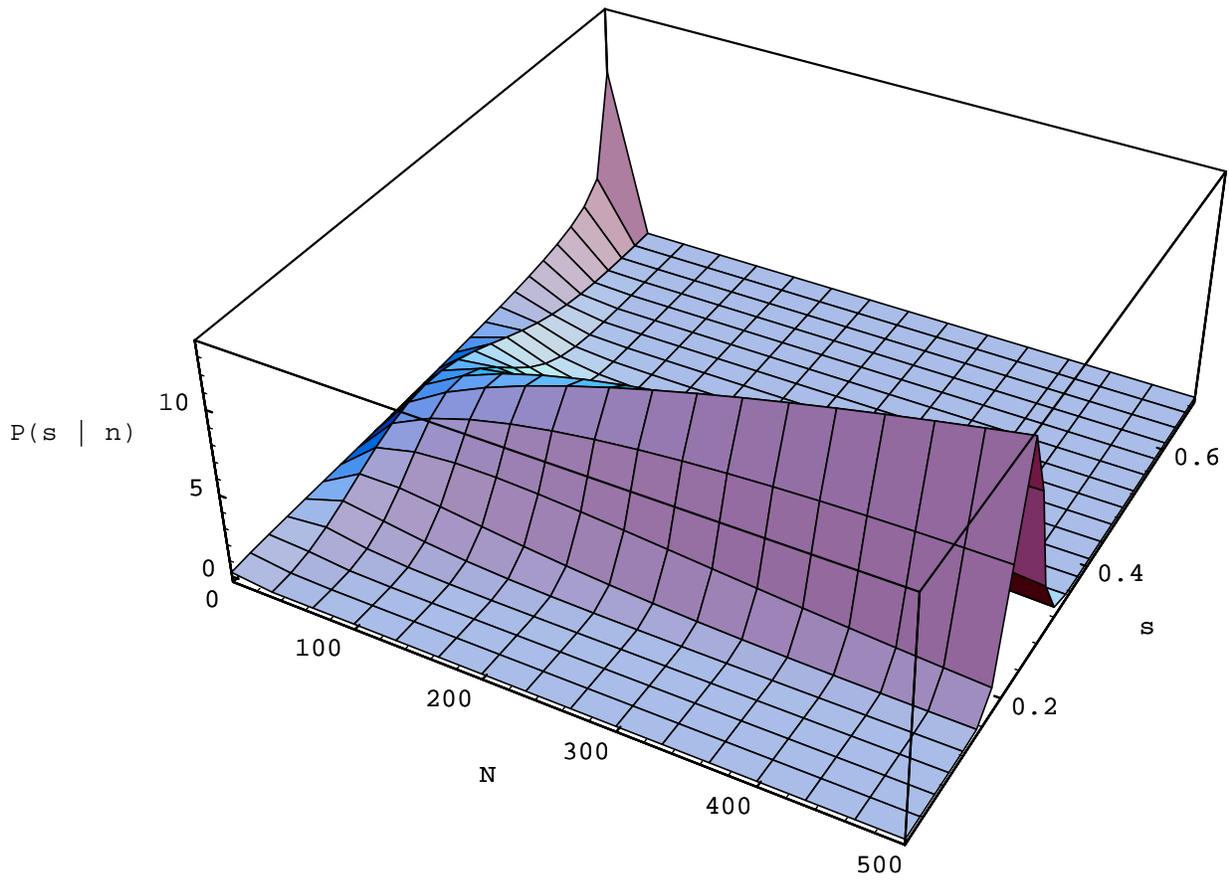

Fig. 6

**FIGURE CAPTIONS**

Figure 1 shows the mean square error (Eqn. 10), for the Bayes (solid) and Frequency Counts (dot-dash) estimators of the entropy $S\left(\mathbf{p}\right) = -\Sigma_{i=1}^{2} p_i \log\left(p_i\right)$.

Figure 2 shows the mean sample variance (Eqn. 11), of the Bayes (solid) and Frequency Counts (dot-dash) estimators of the entropy $S\left(\mathbf{p}\right) = -\Sigma_{i=1}^{2} p_i \log\left(p_i\right)$. Both variances are shown as functions of the sample size N.

Figure 3 shows the sample average (Eqn. 12), for the Bayes (solid) and Frequency Counts (dot-dash) estimators of the two bin (m = 2) entropy $S(\mathbf{p})$. Both are graphed as functions of N for various values of $\mathbf{p} = (p, 1\text{-}p)$. The true value of $S(\mathbf{p})$ is also graphed (dashed).

Figure 4, like figure 3, shows the sample average (Eqn. 12), for the Bayes (solid) and Frequency Counts (dot-dash) estimators of the two bin (m = 2) entropy $S(\mathbf{p})$. However, in figure 4 both are graphed as functions of $\mathbf{p} = (p, 1-p)$ for various values of N. The true value of $S(\mathbf{p})$ is also graphed (dashed).

Figure 5 shows the sample average deviation from true (Eqn. 14), for the Bayes (solid) and Frequency Counts (dot-dash) estimators of the two bin (m = 2) entropy. Both are graphed as functions of $\mathbf{p} = (p, 1\text{-}p)$ for various values of N. The integral over p with the density $P(\mathbf{p})$ appears as figure 1.

Figure 6 shows the posterior pdf (Eqn. 15) of the entropy S($\mathbf{p}$) for m = 2 and fixed counts ratio $n_1 : n_2 = 1 : 15$, but differing overall N = $n_1 + n_2$. As N increases, the density converges to a delta function at the value s = S(1/16 , 15/16) = 0.2338 of the entropy .